\documentclass{PoS}

\title{Supersymmetry with Trilinear R-Parity Violation at the LHC}

\ShortTitle{Supersymmetry with Trilinear R-Parity Violation at the LHC}

\author{\speaker{Nils-Erik Bomark}\\
        University of Bergen, Norway\\
        E-mail: \email{Nils-Erik.Bomark@ift.uib.no}}

\author{Debajyoti Choudhury\\
        University of Dehli, India\\
        E-mail: \email{debajyoti.choudhury@gmail.com}}

\author{Smaragda Lola\\
        University of Patras, Greece\\
        E-mail: \email{Magda.Lola@cern.ch}}

\author{Per Osland\\
        University of Bergen, Norway\\
        E-mail: \email{Per.Osland@ift.uib.no}}

\abstract{The inclusion of trilinear R-parity violating couplings in a supersymmetric theory, has monumental impacts on the phenomenology. Regarding LHC physics, we no longer expect events with a lot of missing transverse energy, but events with large lepton and/or jet multiplicity.

Here we present a study where the standard MSSM scenario of squark and gluino pair production and consequent cascade decay to the neutralino, is followed by a three-body decay of the neutralino through any of the 45 allowed trilinear R-parity violating operators. This allows us to study all these operators and detect any hierarchies among them as well as measuring the neutralino mass.}

\FullConference{The 2011 Europhysics Conference on High Energy Physics, EPS-HEP 2011,\\
		July 21-27, 2011\\
		Grenoble, Rh\^one-Alpes, France}

\begin{document}

\section{Introduction}
What do we hope the LHC will discover?

There are many answers to this question and a large group of them will contain the word supersymmetry.

%
One way of achieving a Supersymmetric scenario very different from the standard one, is to allow R-parity to be violated \cite{barb}. This is done by allowing one or more or the operators,

\begin{equation}\label{Eq:RPV}
    \lambda_{ijk} L_iL_j\bar{E}_k+\lambda'_{ijk}L_iQ_j\bar{D}_k+\lambda''_{ijk}\bar{U}_i\bar{D}_j\bar{D}_k+\mu_iHL_i,
\end{equation}
where $L_i,Q_i,H$ are lefthanded lepton, quark and Higgs superfields and  $E_i,D_i,U_i$ are righthanded lepton, down quark and up quark superfields, to be non-zero. The bilinear couplings, $\mu_iHL_i$, will not be discussed here.

The result of violating R-parity, is a fundamentally different phenomenology as compared to the MSSM, in that all sparticles decay. At first glance it might seem like we lost our dark matter, but there is a viable escape in the form of the gravitino \cite{GravitinoDM}.

For the LHC, the searches for large amounts of missing $E_T$ will have to be replaced with searches for multi-lepton and/or multi-jet final states~\cite{DR}.

Although the couplings of Eq.~(\ref{Eq:RPV}) do allow for a large variety of exotic phenomena, like single sparticle production \cite{single_prod}, direct decay of squarks and sleptons as well as charged particles as end stages of cascade decays (possible since the end stage is no longer expected to be dark matter), it is likely that pair-production still dominates and that the produced particles cascade down to the neutralino which is the lightest sparticle apart from the gravitino \cite{Bomark:2011ye}.

The novel feature, though, is that the neutralino will decay into standard model particles and these decay products will in fact dominate the final state of the event. Since the neutralino couples to all matter fields, this scenario allows us to simultaneously study all the 45 trilinear R-parity violating operators of Eq.~(\ref{Eq:RPV}).

The study discussed here \cite{Bomark:2011ye} is based on Monte Carlo studies of neutralino decays through all 45 operators using the Monte Carlo generator PYTHIA 8 \cite{Sjostrand:2007gs}.

\section{$LL\bar E$}
The large lepton multiplicity of the final state from $LL\bar E$ operators, makes these operators relatively easy to handle. Suppressing the background is more or less trivial, we only require a large number of isolated hard leptons.

The only slight complication arises due to the presence of neutrinos in the final state of the decay. This leads to a loss of information or more precisely momentum, that makes reconstructing the neutralino more difficult. The way around this is to calculate the theoretically expected invariant mass distributions for the charged leptons of the final state. This can be done both for lepton pairs directly from the neutralino as well as lepton pairs where one or both of the leptons come from a decaying tau. We can also calculate the distributions expected for combinations of tau-jets and leptons.

The biggest uncertainty in these calculations concerns the event selection used to collect the data, since we are looking for hard leptons, the low end of the invariant mass distributions will be suppressed, especially for the softer signals including leptons from leptonically decaying taus. It is possible to make a more or less ad hoc suppression of the low end of the distributions that renders reasonable agreement between theory and Monte Carlo data, but one should keep in mind that there are significant uncertainties here.

By fitting the theoretical curves to the data we can now determine which decay channels are open for the neutralino and hence which operators are non-zero. A bonus is that the neutralino mass comes out as one of the parameters of the fit.

\section{$LQ\bar D$}
For $LQ\bar D$ operators, the neutralino normally decays to a lepton plus two jets or a neutrino plus two jets. Due to the smaller amount of leptons as compared to the $LL\bar E$ operators, we here have to put some more care into suppressing backgrounds, especially the $t\bar t$ background. This can be done by requiring two same-sign leptons in the event, this suppresses our signal quite a bit but since lepton pairs from $t\bar t$ events usually have opposite charge, it practically removes the background.

The channel with a charged lepton plus two jets is very promising due to the fact the we can observe everything and therefore we expect to see a peak in the appropriate invariant mass distribution. Unfortunately, this is not always the case, if the lepton in the operator is a tau, the loss of energy due to the neutrino(s) in the decay of the tau, will smear out the peak and thereby make the determination of the neutralino decay channel much more difficult.

If we are really lucky the operator might include b quarks, this would allow easier background suppression due to the possibility for b tagging and if the charge of the b-jet can be measured, one can use same-sign subtraction to reduce combinatorial background in the invariant mass distributions. With b quarks present, also operators with tau leptons can be handled.

The above discussion about b quarks only includes operators with a $\bar D_3$ component, the other possibility with b flavour in the operator is in fact the most difficult case of all $LQ\bar D$ operators. That is the case with $LQ_3\bar D$, i.e.\ a third generation $Q$ operator. The problem here is that the channel that includes a charged lepton in the decay also includes a top quark and therefore will be suppressed by the top mass. The result is that all neutralinos will decay to a neutrino plus two jets, at least one of which is a b-jet. Despite the presence of one or more b-jets, it would be comparatively difficult to firmly establish the presence of this decay channel.

\section{$\bar U\bar D\bar D$}
The most difficult operators are clearly the $\bar U\bar D\bar D$ operators, the reason being that the neutralino typically decays to three jets and we might be left with just a lot of jets, unless the cascade decay chain gives us something more useful. A lot of jets is very difficult to separate from QCD background but this has been proven possible \cite{Butterworth:2009qa}.

One special group of $\bar U\bar D\bar D$ operators are the ones with top flavour, i.e.\ the three operators $\bar U_3\bar D\bar D$. If the neutralino is lighter than the top quark, this may lead to a scenario where the neutralino has no simple decay channels and can leave the detector, in other words, we may be facing a fake MSSM scenario.

The other possibility, that the neutralino is in fact heavier than the top, would lead to neutralino decays to a top (or anti-top) plus two jets. Since the whole event contains two neutralinos, in half the cases we would see events with two top quarks or two anti-top quarks. The easiest way to extract these events is with the two same-sign lepton criteria used for the $LQ\bar D$ operators.

\section{Summary and conclusions}
Among all possible types of new physics searched for at the LHC, supersymmetry sticks out as somewhat of a favorite among both theoreticians and experimentalists. One interesting possibility within supersymmetry is to violate R-parity, which in a large part of parameter space yields neutralino decays to three standard-model fermions. This decay offers the possibility to study all trilinear R-parity violating operators simultaneously.

In this presentation, the possibility to detect such scenarios and to measure which R-parity violating  couplings are large, were discussed. It has been shown that for most couplings the prospects of successfully determining the coupling type and flavours are very good. This is true also if more than one operator is large.

Some more difficult cases include $L_3Q\bar D$ type operators where the tau in the final state makes identification more difficult, $LQ_3\bar D$ type operators where the large top mass suppresses the decay channel including charged leptons and some $\bar U\bar D\bar D$ operators where the purely hadronic final state can make the QCD background difficult to suppress. There is also a risk that $\bar U_3\bar D\bar D$ operators yield a scenario looking exactly like the R-parity conserving MSSM, but where the neutralino decays outside the detector.


\end{document}